# Use of Superparamagnetic Nanoparticle/Block Copolymer Electrostatic Complexes as Contrast Agents in Magnetic Resonance Imaging

**Jean-François Berret**
Matière et Systèmes Complexes, UMR 7057 CNRS Université Denis Diderot Paris-VII,
140 rue de Lourmel, F-75015 Paris France
jean-francois.berret@paris7.jussieu.fr

**Régis Cartier***
Klinik für Anästhesiologie und Operative Intensivmedezin, Univeritätsmedezin Berlin,
Spandauer Dam 130, D-14050 Berlin, Germany

**Abstract** : During the past years we have investigated the complexation between nanocolloids and oppositely charged polymers. The nanocolloids examined were ionic surfactant micelles and inorganic oxide nanoparticles. For the polymers, we used homopolyelectrolytes and block copolymers with linear and comb architectures. In general, the attractive interactions between oppositely charged species are strong and as such, the simple mixing of solutions containing dispersed constituents yield to a precipitation, or to a phase separation. We have developed means to control the electrostatically-driven attractions and to preserve the stability of the mixed solution. With these approaches, we designed novel core-shell nanostructures, e.g. as those obtained with polymers and iron oxide superparamagnetic nanoparticles. In this presentation, we show that electrostatic complexation can be used to tailor new functionalized nanoparticles and we provide examples related to biomedical applications in the domain of contrast agents for Magnetic Resonance Imaging.

## Introduction

Magnetic nanoparticles are currently used in a wide variety of material science and biomedical applications. Important technological advances have been achieved in the purification of biomolecules and in cell separation techniques. Surface-modified nanoparticles are also developed for *in vivo* applications through highly specific Magnetic Resonance Imaging (MRI) and drug delivery systems. During the last years, the functionalization of nanoparticles has become a key issue to permit a controlled assembly of nanostructures with a defined molecular architecture and specific physicochemical properties. The work described in this extended abstract aims to develop and characterize self-assembled functional magnetic nanocolloids with controlled hierarchical morphologies.

A broad range of techniques in chemistry and physical chemistry are now being developed for the coating of inorganic nanoparticles [1]. The techniques of interest are the adsorption of charged ligands or stabilizers on their surfaces [2-4], the layer-by-layer deposition of polyelectrolyte chains [5] and the surface-initiated polymerization resulting in high-density polymer brushes [6]. Other approaches have focused on the encapsulation of the particles in amphiphilic block copolymer micelles [7,8].

In the present communication, we are following a strategy based on the principle of electrostatic complexation. The nanoparticles were complexed using asymmetric block copolymers, where one block is of opposite charge to that of the particles and the second block is neutral. Recently, it has been shown that polyelectrolyte-neutral copolymers can associate in aqueous solutions with oppositely charged surfactants [9-13], macromolecules [14-17] and proteins [18,19] and are capable of building stable "supermicellar" aggregates with core-shell structures. The core of radius ~ 20 nm can be described as a dense coacervate microphase comprising the oppositely charged species. The corona of thickness ~ 20 – 50 nm is made by the neutral chains and ensures the colloidal stability of the whole.

In this paper, we have shown that it was possible to use electrostatic self-assembly in order to stabilize and associate $\gamma$-$Fe_2O_3$ superparamagnetic nanometer size particles in a controlled manner. As anticipated, we demonstrated that the core of the mixed aggregates is made of densely packed nanoparticles. For the first time, the aggregation numbers (*i.e.* the number of particles per magnetic core) could be estimated directly from the cryo-TEM pictures. Magnetic resonance spin-echo measurements indicated a significant increase of the ratio between the transverse and longitudinal relaxivities, which usually tests the efficiency of contrast agents in MRI. This result suggested that the polymer-nanoparticle hybrids designed by this technique could be used as $T_2$-contrast agents for biomedical applications.

## Experimental

The cationic-neutral diblocks were synthesized by controlled radical polymerization according to MADIX technology [20]. The polyelectrolyte portion is a positively charged poly(trimethylammonium ethylacrylate) block, whereas the neutral portion is poly(acrylamide) [11,12]. Three molecular weights were put under scrutiny, corresponding to 7 (2 000 g·mol$^{-1}$), 19 (5 000 g·mol$^{-1}$) and 41 (11 000 g·mol$^{-1}$) monomers in the charged blocks and 420 or 840 (30 000 g·mol$^{-1}$ or 60 000 g·mol$^{-1}$) for the neutral chain. In the following, the copolymers are abbreviated as PTEA$_{2K}$-b-PAM$_{60K}$, PTEA$_{5K}$-b-PAM$_{30K}$ and PTEA$_{11K}$-b-PAM$_{30K}$. The role of the neutral chains was to prevent the coacervate microphase to reach micron sizes at the mixing that would finally settle down over time. The superparamagnetic nanoparticles of maghemite ($\gamma$-Fe$_2$O$_3$) were made available to us by the Laboratoire des Liquides Ioniques et Interfaces Chargées, Université Pierre et Marie Curie (Paris, France). The particles were characterized using different techniques, such as vibrating sample magnetometry, scattering and cryogenic electron microscopy experiments. This size distribution of the particles was found to be well-accounted for by a log-normal function, with a most probable diameter 6.3 nm and polydispersity 0.23. At the pH values at which the complexation occurred (pH 7 – 8), the particles were stabilized by electrostatic interactions mediated by charged ligands. The ligands adsorbed on the surface of the particles were citric acid in its sodium salted form (sodium citrate). Polymer-nanoparticle complexes were obtained by simple mixing of stock solutions prepared at the same weight concentration and pH. The relative amount of each component is monitored by the mixing ratio X, which is defined as the ratios of the volumes of nanoparticle solution added relative to the polymer solution. Protocols for mixing oppositely charged species in solutions have been described previously [21].

**Light Scattering and Cryogenic Electron Microscopy**

Dynamic light scattering performed on $\gamma$-Fe$_2$O$_3$|PTEA$_{5K}$-b-PAM$_{30K}$ solutions at c = 0.2 wt. % reveals the presence of purely diffusive relaxation modes for all values of X. Fig. 1 displays the evolution of the average hydrodynamic diameter $D_H$ derived from the time dependence of the autocorrelation functions and from the Stokes-Einstein relationship. For X > 0.01, $D_H$ ranges between 60 nm and 80 nm. At large X (X > 5), a second mode associated to the single nanoparticle became apparent. In this range, the autocorrelation is fitted by a double exponential decay. $D_H$-values as large as 60 – 80 nm are well above those of the individual components of the system. The polydispersity index resulting from the cumulant analysis is found within a range of 0.10 to 0.25. In the inset of Fig. 1, cryo-TEM images illustrate the core-shell microstructure of the mixed aggregates. The photographs cover spatial fields that are approximately 0.2×0.3 μm$^2$ and display clusters of nanoparticles. For contrast reasons, only the inorganic cores are visible with this technique. The extension of the polymer corona is shown by a circle of diameter $D_H$.

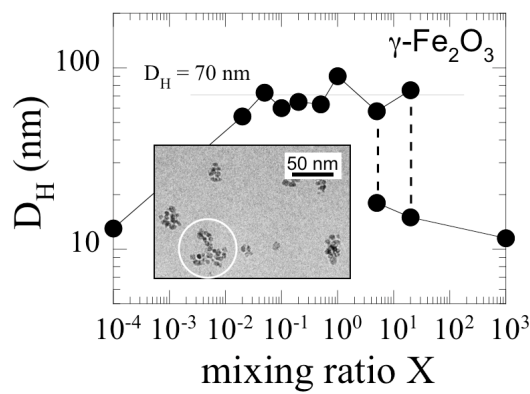

**Figure 1** : Hydrodynamic diameter $D_H$ as a function of the mixing ratio X for $\gamma$-Fe$_2$O$_3$|PTEA$_{5K}$-b-PAM$_{30K}$ solutions. Inset : cryo-TEM image of the nanoparticle clusters.

**Relaxometry**

Figs. 2a and 2b show the inverse relaxation times $1/T_1$ and $1/T_2$ as a function of the iron molar concentration [Fe] for the citrate-coated nanoparticles and for the nanoparticle-copolymer aggregates [21,22]. Samples were prepared at X = 1 for PTEA$_{5K}$-b-PAM$_{30K}$ and at X = 2 for PTEA$_{11K}$-b-PAM$_{30K}$, both at concentration c = 0.2 wt. %. In order to monitor the inversion-recovery and spin echo pulse sequences, the solutions were diluted by a factor of 10 to 1000. For these experiments, light scattering was used to verify that the size and microstructure of the polymer-nanoparticle complexes were not modified under dilution. In Figs. 2a and 2b, the inverse relaxation times were found to vary linearly with the iron concentration, according to the following equation [23] :

$$\frac{1}{T_{1,2}([Fe])} = R_{1,2} \cdot [Fe] + \frac{1}{T^0_{1,2}} \qquad (1)$$

where $R_1$ and $R_2$ are the longitudinal and transverse relaxivities, respectively. The intercepts $1/T^0_{1,2}$ are the proton inverse relaxation times in pure water. At a Larmor frequency of 20 MHz, we found $T^0_1$ = 4.0 s and $T^0_2$ = 1.6 s. The data in Fig. 2a shows that with increasing cluster size, the longitudinal relaxivity $R_1$ exhibits a slight decrease, but remains around 20 s$^{-1}$·mM$^{-1}$. The transverse relaxivity $R_2$ on the contrary increases noticeably. $R_2$ starts at 39 ± 2 s$^{-1}$·mM$^{-1}$ for the bare nanoparticles, rises up to 74 ± 4 for clusters prepared with the 5K-30K copolymer, and culminates at 162 ± 4 for clusters made with the 11K-30K chains. Such values for $R_1$ and $R_2$ suggest that the polymer-nanoparticle aggregates created by electrostatic self-assembly could serve as $T_2$ contrast agents [21,22]. The results reported in Figs. 2 support recent studies by Martina *et al*. [24], Manuel-Perez *et al*. [25] and Roch and coworkers [26]. These reports all indicate that clusters of magnetic nanoparticles have enhanced transverse relaxivities $R_2$, and relaxivity ratios $R_2/R_1$.

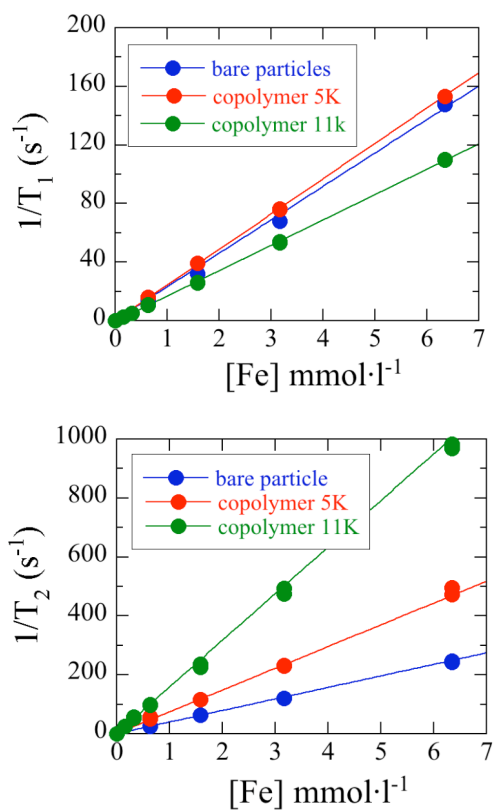

*Figure 2 :* Inverse longitudinal and transverse relaxation times $1/T_1$ (a) and $1/T_2$ (b) for mixed copolymer-particle hybrids as a function of the iron molar concentration [Fe]. The straight lines were calculated according to Eq. 1. The solutions were prepared at the ratio where all the components present in solution associate to form complexes.

**Magnetic resonance Imaging**

In order to assess the role of the neutral corona in *in vivo* conditions, dilute solutions of γ-$Fe_2O_3$│$PTEA_{5K}$-*b*-$PAM_{30K}$ were injected intravenously to 300 g Wistar rats (Harlan Winkelmann, Borchen, Germany). The visualization of the liver was monitored by spatially resolved resonance magnetic imaging performed in a 3 T magnetic resonance scanner (Sigma 3T94, General Electric Healthcare, Milwaukee, WI) [27]. The dose of coated injected nanoparticles were 0.6 mg and 1.2 mg of iron per kilogram of body weight. The results obtained with the electrostatic complexes were compared with a commercially available standard (Resovist, carboxydextran coated iron oxide nanoparticles). The $T_2$-weighted images of the liver were recorded before injection, 15 minutes and 60 minutes after injection. As illustrated in Fig. 3, at 15 mn post-injection the liver has darkened considerably for both Resovist and coated nanoparticle clusters. The contrast enhancement of the liver in the $T_2$-weighted sequences indicated that there was accumulation of magnetic material in the liver. In other term, the neutral polyacrylamide corona was not sufficient to avoid, or even to slow down the capture and removal of the contrast agents by the immune system of the rat. Finally, it is important to recall that in MRI the measured signal does not originate from the contrast agent only. Instead, the technique measured the signal from the tissue (relaxation time of water protons in the tissue) which might be altered by the presence of the contrast agent. This is the reason why the liver appeared grey in the native scans. $T_2$-relaxations are prolonged by the presence of magnetic inhomogeneities caused by the particles, generating additional contrast to the environment. Therefore, the signal at least depends on the tissue signal and on the magnetic properties of the contrast agent. The darkness of the final image also depends on how the material is distributed in the tissue. More work is needed to quantify the iron oxide concentration in the liver as a function of time.

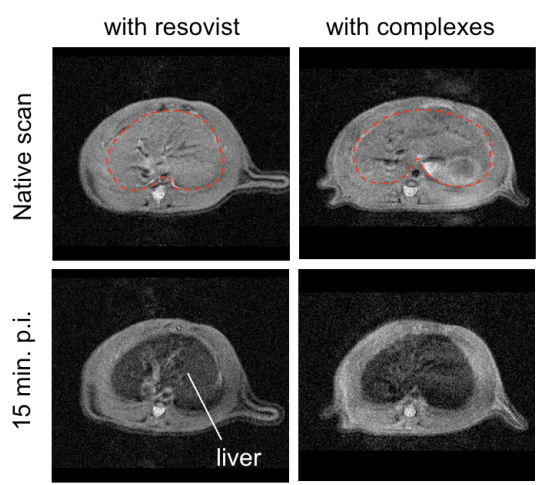

*Figure 3 :* Magnetic resonance signal intensity of liver of rats in $T_2$-weighted sequences. Images were acquired before and 15 min after intravenous injection of nanoparticles in a concentration of 1.2 mg of Fe/kg of body weight.

*Conclusion*

In this paper, we investigated the formation of new hybrid aggregates comprising charged copolymers and nanoparticles. The mechanism at the origin of the self-assembly is based on electrostatic adsorption and charge compensation between oppositely charged species. The systems put under scrutiny were cationic-neutral hydrosoluble block copolymers and 6.3 nm diameter superparamagnetic γ-$Fe_2O_3$ nanoparticles. The properties of the nanoparticles clusters coated with a neutral corona were tested in relaxometry and in *in vivo* magnetic resonance imaging experiments. A significant result is that the transverse relaxivity, $R_2$ is noticeably increased with the size of the magnetic clusters. The ratio $R_2/R_1$, which is usually an important parameter in estimating the efficiency between $T_2$-contrast agents, increases from 1.7 for the citrate-coated nanoparticles to 9.3 for $\gamma\text{-Fe}_2\text{O}_3|\text{PTEA}_{11K}\text{-}b\text{-PAM}_{30K}$. $T_2$-weighted sequences on rat livers have however indicated that after intravenous injection, these clusters are filtrated by the liver as rapidly as a commercial product.